\documentclass[useAMS,usenatbib]{mn2e}

\usepackage{graphicx}
\usepackage{epsfig}

\title[The Evolution Of Carbon, Sulphur, and Titanium Isotopes]{The Evolution 
Of Carbon, Sulphur, and Titanium Isotopes from High-Redshift to the Local 
Universe}
\author[G.~L. Hughes et~al.]{G.~L.~Hughes$^1$, B.~K.~Gibson$^1$, 
L.~Carigi$^{1,2}$, P.~S\'anchez-Bl\'azquez$^1$, J.~M.~Chavez$^{3}$\newauthor 
and D.~L.~Lambert$^{3}$ \\
$^1$University Of Central Lancashire, Centre for Astrophysics, Preston, PR1~2HE, United Kingdom\\ $^2$Instituto de Astronom\'ia, Universidad Nacional Aut\'onoma de M\'exico, A.P. 70-264, M\'exico, 04510 D.F., Mexico\\ $^3$University Of Texas, Department of Astronomy, Austin, Texas, 78712, USA}
\date{Released 2008 Xxxxx XX}

\pagerange{\pageref{firstpage}--\pageref{lastpage}} \pubyear{2008}

\def\LaTeX{L\kern-.36em\raise.3ex\hbox{a}\kern-.15em
    T\kern-.1667em\lower.7ex\hbox{E}\kern-.125emX}

\begin{document}

\label{firstpage}

\maketitle

\begin{abstract}
Recent observations of carbon, sulphur, and titanium isotopes at redshifts 
z$\sim$1 and in the local stellar disc and halo have opened a new window
into the study of isotopic abundance patterns and the origin of the 
chemical elements.  Using our Galactic chemical evolution code {\tt GEtool}, 
we have examined the evolution of these isotopes within the framework
of a Milky Way-like system.
We have three aims in this work: first, to test the claim that novae are 
required, in order to explain the carbon isotope patterns in the Milky Way;
second, to test the claim that sulphur isotope patterns at high-redshift 
require an initial mass function biased towards massive stars; and third, 
to test extant chemical evolution models against new observations of
titanium isotopes that suggest an anti-correlation between 
trace-to-dominant isotopes with metallicity.
Based upon our dual-infall galactic chemical evolution modelling of 
a Milky Way-like system, and the subsequent comparison with these new and 
unique datasets, we conclude the following: novae are not required to 
understand the evolution of $^{12}$C/$^{13}$C in the solar neighbourhood; 
a massive star-biased initial mass function is consistent with the low 
ratios of $^{12}$C/$^{13}$C and $^{32}$S/$^{34}$S seen in one 
high-redshift late-type spiral, but the consequent super-solar metallicity 
prediction for the interstellar medium in this system seems highly 
unlikely; and deficient isotopes of titanium are predicted to correlate 
positively with metallicity, in apparent disagreement with the new 
datasets; if confirmed, classical chemical evolution models of the Milky Way
(and the associated supernovae nucleosynthetic yields)
may need a substantial overhaul to be made consistent.
\end{abstract}


\section{Introduction}
Galactic chemical evolution models are employed to study the spatial and 
temporal evolution of elements and isotopes throughout the Universe. When 
coupled to phenomenological representations of galaxy assembly, such 
models can be compared directly with exquisite elemental and isotopic 
abundance patterns observed locally in the Milky Way. From these comparisons, 
conclusions can be drawn regarding the veracity of the underlying 
micro-physics governing stellar evolution and nucleosynthesis, in addition 
to the macro-physics governing the assembly of galaxies, the redistribution 
of the interstellar medium (ISM) over galactic timescales, and the 
relative birthrate of stars of various masses (the so-called initial mass 
function, or IMF).  While most galactic chemical evolution models to 
date have concentrated on predictions related to the total elemental 
abundance patterns, in unique circumstances, the availability of 
detailed isotopic patterns can enhance the predictive power of these
models; such isotopic patterns afford additional leverage for
discriminating between the various origin sites for the chemical 
elements.  A full literature review of the field would be 
unwieldy, but we refer the reader to any number of comprehensive 
reviews and the many important references therein - 
e.g. Timmes, Woosley \& Weaver (1995, hereafter TWW95), 
Prantzos, Aubert \& Audouze (1996), Fenner \& 
Gibson (2003), Romano \& Matteucci (2003, hereafter, RM03), and 
Chiappini et~al. (2008).

Recent observational work has opened a new window into the study of 
isotopic abundance patterns - specifically, the identification (for the 
first time) of carbon and sulphur isotopes at redshifts z$\sim$1 
(Muller et~al. 2006, hereafter M06; Levshakov et~al. 2006), coupled with the 
recent determination of titanium isotopic abundances in the local stellar 
disc and halo (Chavez 2008). The abundances of each of these isotopes, and 
their evolution with redshift, hold clues as to the relative importance 
of supernovae versus asymptotic giant branch stars versus novae in seeding 
the Universe with these important elements.

In this work, we present predictions for isotopic ratios of 
carbon, sulphur, and titanium within the framework of a classical 
Milky-Way-like disc galaxy model. The paper is organised as follows: the 
fundamental nucleosynthetic origins of the relevant
carbon, sulphur, and titanium isotopes are first (briefly) reviewed
in \S~2; in \S~3, we introduce the chemical evolution code ({\tt GEtool}) 
and the four stellar yield compilations employed in our modelling;
finally, the results are presented and summarised in 
\S~4 and \S~5, respectively.

\section{Origin Of Carbon, Sulphur, and Titanium Isotopes}
Because we in the astronomical community are generally more 
accustomed to discussing \it elemental \rm abundance patterns, 
rather than \it isotopic \rm patterns, we felt it would be 
useful to provide an overview of the nucleosynthetic origins of
these relevant isotopes.
Much of this section has been derived and summarised from Clayton's
(2007) exceptional handbook, to which the reader is referred 
for definitive and comprehensive descriptions.
Complementary 
discussions of the relevant nucleosynthesis processes and their products
can be found in Woosley \& Weaver (1995),
Arnett (1996), Pagel (1997), and Matteucci (2001).

\subsection{Carbon-12}
As the initial product of helium burning (the classical
triple-$\alpha$ process), $^{12}$C is the second most 
abundant nucleus formed by nucleosynthesis in stars. While
the exact accounting remains uncertain, significant carbon
production can likely be traced to massive stars that
eventually become Type~II supernovae (e.g.  Chiappini et~al. 1997),
with a substantial contribution also derived
from intermediate mass asymptotic giant branch (AGB) stars
(e.g. Carigi et~al. 2005).
The newly created $^{12}$C nuclei in these latter 
stars is convected to the surface, 
often leading to the formation of a carbon star; ultimately,
this carbon-enriched envelope is lost to the ISM through the process of
stellar winds associated with planetary nebulae.  The amount of $^{12}$C
produced depends critically upon the $^{12}$C + $^4$He 
$\rightarrow ^{16}$O reaction.

\subsection{Carbon-13}
The $^{13}$C isotope is considered a secondary nucleus, produced not from
the nuclear fusion of hydrogen and helium, but as a secondary process
involving ``seed'' nuclei of $^{12}$C.  It is thought to originate within
stars not massive enough to become supernovae, in particular, AGB stars.
At sufficient temperatures, $^{13}$C is produced via the capture of a 
proton by the $^{12}$C nucleus, to form $^{13}$N, which itself 
undergoes $\beta$-decay to form $^{13}$C. The relative abundance
depends upon both the relevant reaction and destruction rates.

Having said that, there is also evidence to suggest that $^{13}$C 
nucleosynthesis also has a primary component (i.e., a production pathway 
exists which does not depend upon seed nuclei of $^{12}$C).  For example, 
within AGB stars, periodic dredge-up episodes bring newly-formed 
$^{12}$C to the surface, converting AGB stars to carbon stars, as noted 
above.  {\it If } the temperature at the base of the envelope is sufficient, this 
primary $^{12}$C can be partly converted to (primary) $^{13}$C and 
$^{14}$N by the first two reactions of the CN-cycle.  However, if the star is 
massive enough (M $>$ 4M$_{\odot}$) then hot bottom burning can occur, 
delaying or preventing the AGB star from turning into a carbon star, and therefore 
this process.

\subsection{Sulphur-32}
$^{32}$S is formed mostly through oxygen burning, two $^{16}$O nuclei 
colliding to form $^{28}$Si and $^4$He, with these products subsequently
fusing to yield $^{32}$S.  Almost all $^{32}$S is 
produced in Type~II supernovae, which eject approximately 
ten times the quantity synthesised within Type~Ia supernovae, and
occur roughly fives times as often.

\subsection{Sulphur-34}
$^{34}$S originates as a byproduct of oxygen-burning. $^{34}$S is 
partly a secondary isotope as it is formed from newly-made $^{32}$S and 
$^{33}$S by neutron captures, which itself is aided if the star that creates 
them also has carbon and oxygen in its composition. $^{18}$O and 
$^{22}$Ne created from this initial carbon and oxygen produce extra 
neutrons that are needed by heavier sulphur isotopes, but excess neutrons 
are also produced during oxygen burning by positron emissions, and as 
such, $^{34}$S is also partly a primary isotope. As for
$^{32}$S, $^{34}$S is produced primarily from supernovae.

\subsection{Titanium-46}
$^{46}$Ti originates from oxygen- and silicon-burning in massive stars. 
Two $^{16}$O nuclei collide and subsequent $\alpha$-captures produce 
$^{44}$Ti, while the capture of two free neutrons results in $^{46}$Ti. 
This can also be viewed as the addition of an $\alpha$-particle to 
$^{42}$Ca, so $^{46}$Ti becomes abundant in the same oxygen-burning zone 
that synthesises $^{42}$Ca.
If the burning continues into that of silicon, then the $^{46}$Ti 
abundance erodes quickly. $^{46}$Ti cannot be labelled as a secondary nucleus 
as there are positive $\beta$-decays during the oxygen burning.

\subsection{Titanium-47}
$^{47}$Ti also originates from oxygen- and silicon-burning in massive stars, 
but in this case, three free neutrons are captured. This can be viewed as 
the addition of a neutron to $^{46}$Ti, so $^{47}$Ti becomes abundant in 
the same oxygen-burning zone that synthesises $^{46}$Ti.
Some models of Type~Ia supernovae also contribute to 
interstellar $^{47}$Ti. There does appear to be an apparent problem though, 
in the sense that models of both supernovae types have been claimed to
be insufficient producers of $^{47}$Ti with respect to observations
(Timmes et~al. 1995).

\subsection{Titanium-48}
The production of $^{48}$Ti is traced to the nucleosynthesis of $^{48}$Cr 
in stellar explosions; two 
subsequent $\beta$-decays after its ejection leads to 
$^{48}$Ti. This occurs mostly in explosive silicon-burning and during helium 
fusion. The latter could occur either by the $\alpha$-rich freezeout of 
shock-decomposed nuclei near the core of a Type~II supernova, or as
part of explosive helium burning associated with Type~Ia supernovae.

\subsection{Titanium-49}
$^{49}$Ti is produced mainly by the nucleosynthesis of radioactive 
$^{49}$Cr in stellar explosions of both types of supernovae. The isotope 
$^{49}$Cr is the result of the explosive fusion of helium as outlined above; 
$^{49}$Cr is also synthesised during silicon-burning, as for the lighter 
titanium isotopes.

\subsection{Titanium-50}
It has been suggested that $^{50}$Ti is produced primarily in sub-Chandrasekhar
mass Type~Ia supernovae (Timmes et~al. 1995), 
during which electron capture turns the 
composition neutron-rich.  Some fraction of the $^{50}$Ti is likely
also made by slow neutron capture within the the burning shells of
pre-supernovae massive stars and AGB stars.  It would appear that 
explosive burning in Type~II supernovae does not produce 
significant quantities of $^{50}$Ti.

\section{The Chemical Evolution Model}
Throughout our work, we have used the {\tt GEtool} (Fenner \&
Gibson 2003; Fenner, Murphy \& Gibson
2005) galactic chemical evolution package, employing its default
``dual infall'' (halo + disc) mode (similar in spirit to the 
seminal models of Chiappini, Matteucci \& Gratton 1997).

Within this framework, the halo phase occurs on a rapid timescale and 
enriches the initially primordial gas to a metallicity of $\sim$10\% solar. The 
second (disc) phase is delayed by $\sim$1~Gyr with respect to the first, and
acts over a more prolonged timescale.  We assume the infall of fresh material 
during this second phase to be somewhat metal-enriched 
(10\% solar) and $\alpha$-enhanced (0.4~dex), consistent with 
patterns seen in metal-poor halo/thick disc
stars (e.g. Ryan, Norris \& Beers 1996) 
and present-day high-velocity infalling halo gas (e.g. Gibson et~al. 2001).

The rate at which material is accreted is assumed to decline exponentially. 
The evolution of total surface mass density $\sigma_{tot}(r, t)$ is given by: 

\begin{equation}
\frac{d\sigma_{tot}(r, t)}{dt} = A(r)e^{-t/\tau_H} + B(r)e^{-(t-t_{delay})/\tau_D(r)} 
\end{equation}

\noindent
where the infall rate coefficients $A(r)$ and $B(r)$ are chosen in order to 
reproduce the present-day surface mass density of the halo and disk 
components, which we take to be 10 and 45 $M_{\odot}$ pc$^{-2}$, 
respectively. The adopted timescales for the infall phases in
the solar neighbourhood ($r_{\odot}$=8.5kpc) 
are $\tau_H =$ 0.05~Gyr and $\tau_D =$ 10.5~Gyr,
with the functional form for the latter being 
$\tau_D(r)$=$1.38r-1.27$, reflecting the ``inside-out'' formation framework in 
which the dual-infall model operates. These timescales and coefficients provide 
model predictions consistent with various local observational 
constraints such as the metallicity distribution function, 
age metallicity relation, and present-day gas 
surface density distribution, and are consistent with those
employed by Fenner et~al. (2005).

We adopt a fairly conservative star formation prescription based upon
a ``Schmidt Law'' of the form:

\begin{equation}
\psi(r, t) = \nu\sigma^2_{gas}(r, t)
\end{equation}
 
\noindent
where the value of the star formation efficiency $\nu$ is constrained by 
the present-day gas fraction (for this work, $\nu$$=$0.06~Gyr$^{-1}$).

\subsection{Initial Mass Function}
The shape of the initial mass function (IMF) controls the 
fraction of material locked-up in stellar generation, which in turn
determines the rate at which different elements are released into the
ISM.  Our default assumption is that of the three-component IMF of 
Kroupa, Tout \& Gilmore (1993; hereafter, KTG), with lower- and upper-mass
limits of 0.08 and 60~M$_{\odot}$, respectively; the KTG IMF lies between 
those of Salpeter (1955) and Scalo (1986), in terms of mass fraction 
tied up in Type 
II supernovae progenitors (e.g. Table~7 of Gibson 1997).  Unless
otherwise stated, we assume that the mass fraction of the IMF which 
is tied up in SNe~Ia progenitor binary systems (total binary
masses in the range 3$-$16~M$_\odot$) is 4\%; such an assumption yields,
within our adopted model formalism,
a disc-averaged present-day ratio between SNe~II and SNe~Ia rates of
4.1 (consistent with the disc-averaged empirical SNe~II to SNe~Ia
ratio of 3.7 -- van~den~Bergh 1988).
We supplement this by exploring a range of single power-law IMF slopes
to isolate the relative contributions of low- and high-mass stars.

\subsection{Stellar Yields}
In order to sample the range of uncertainties inherent to stellar 
evolution modelling, we explore the use of several sets of 
metallicity-dependent nucleosynthetic yields, in this work --
those of Woosley \& Weaver (1995), Chieffi \& Limongi (2004), 
Kobayashi et~al (2006),\footnote{Kobayashi et~al. (2006) provide 
yields for both Type~II supernovae {\it and} hypernovae, the latter
represented by explosion energies 10$\times$ those of their
supernovae models.}
for Type~II supernovae,\footnote{None of the
Type~II supernovae yield compilations employed here take into account
the effects of rotationally-induced mixing; while this has little
effect at moderate-to-high (e.g. Galactic disc) metallicities, 
it may have a significant
impact at low (e.g. galactic halo) metallicities (Meynet et~al. 2006;
Hirschi 2007).}
and those of Karakas \& Lattanzio (2007), for low- and intermediate-mass 
single stars (hereafter, WW95, CL04, K06, and KL07, respectively). For 
Type~Ia supernovae, the yields of Nomoto et~al. (1997) have been 
assumed.

As these yields only have data for M~$\leq$~40~M$_{\odot}$, and our
{\tt GEtool} models assumed
an upper-mass limit of 60~M$_{\odot}$, a linear extrapolation was 
used, to extend the yields to the highest masses.  We stress though
that the results do not depend upon the assumption of linear, 
as opposed to logarithmic or, indeed, ``flat'' extrapolation.

\begin{table*}
\begin{minipage}{126mm}
\caption{Observational Constraints Used}
\label{tab:data}
\begin{tabular}{@{}lccccc}
\hline
Isotope & Meteoritic Grain$^{a,1}$ & ISM$^{a,2}$ & z=0.89$^{b,3}$ & z=1.15$^{c,4}$  \\
\hline
$^{12}$C/$^{13}$C & 82.83 & 70 $\pm$ 10 & 27 $\pm$ 2 & $>$80 & \\
$^{32}$S/$^{34}$S & - & - & 10 $\pm$ 1 & - \\
\hline
\end{tabular}

\begin{tabular}{@{}lcccccc}
\hline
Isotope & Solar System$^{d,1}$ & Solar System$^{e,1}$ & ISM$^{e,2}$ & $Z\leq10^{-3}$$^{f,5,*}$ \\
\hline
$^{12}$C/$^{13}$C & 89 & - & - & 19.38 \\
$^{32}$S/$^{34}$S & 22 & 22.64 & 24.4 $\pm$ 5 & - \\
\hline
\end{tabular}

\medskip
$^a$RM03; $^b$M06; $^c$Levshakov et~al. (2006);
$^d$Anders \& Grevesse (1989); $^e$Mauersberger et~al. (2004); $^f$Spite et~al. (2006).
\\\\
$^1$t=8.5~Gyr; $^2$t=13~Gyr; $^3$t=5.7~Gyr; $^4$t=4.8~Gyr; $^5$t$\sim$~1~Gyr; 
redshift-to-age conversion assuming
$\Lambda$CDM concordant cosmology (H$_\circ$=71~km~s$^{-1}$~Mpc$^{-1}$, 
$\Omega_{\rm M}$=0.27, $\Omega_\Lambda$=0.73).
\\\\
\verb+*+ The mean $^{12}$C/$^{13}$C value for the unmixed stars from Spite (2006). 
\end{minipage}
\end{table*}

A detailed isotope-by-isotope comparison between the compilations
is beyond the scope of our work; indeed, 
as ``end-users'' \it only \rm of the products of
the aforementioned sophisticated stellar evolution and nucleosynthesis
codes, one could argue it is not even a feasible undertaking.  What
\it is \rm important though, from our perspective, is that by 
employing a range of yield compilations, we are in some sense
sampling the ``convolution'' of a fair range of stellar ``micro-physics''
uncertainties intrinsic to these codes.  
This is, realistically, the best compromise for 
end-users to take.

As noted earlier, we 
will be concentrating upon the chemical evolution of the primary
isotopes of carbon, sulphur, and titanium, in response to 
recent advances pertaining to their observation locally and in the
distant Universe.
To foreshadow the chemical evolution results described in \S~4, in
Figs.~\ref{fig:c_s.isotopes} and \ref{fig:ti_isotopes} we
show the $^{12}$C/$^{13}$C, $^{32}$S/$^{34}$S, and 
$^{46,47,49,50}$Ti/$^{48}$Ti isotope ratios, as a function of stellar mass 
(at solar metallicity) for each of the yield compilations adopted here.

We can see from Fig.~\ref{fig:c_s.isotopes} that, for carbon, there is 
little difference between supernovae and hypernovae (K06-SNe and
K06-HNe, respectively), whereas for sulphur, the explosion energetics
lead to a factor of $\sim$3 variation in the predicted ejecta ratios.
We will return to the issue of the impact of apparently ``discrepant''
individual stellar models in \S~4, but we should note now, for example, 
the obvious ``outlier'' seen in the K06 M=25~M$_\odot$ solar metallicity
model (upper panel of Fig.~\ref{fig:c_s.isotopes}); this 
two-orders-of-magnitude outlier is also seen in the K06 M=18~M$_\odot$
Z=0.004 (not plotted) model.  In both cases, this can be traced to 
the respective models producing $\sim$100$\times$ the $^{13}$C of 
the ``flanking'' models. In Fig.~\ref{fig:ti_isotopes}, it is also seen that 
there is a factor of $\sim$3 variation in the predicted ejecta ratios of 
K06-SNe and K06-HNe.

Figures~\ref{fig:ww95_metallicity} and \ref{fig:kl07_metallicity} show the 
variation of $^{12}$C/$^{13}$C and $^{32}$S/$^{34}$S, with mass, for 
different metallicities.  For clarity, only the WW95 and KL07 yields are 
shown, as the CL04 and K06 yields demonstrate a similar 
metallicity-dependence as that of WW95; the general trend of decreasing 
isotope ratio with increasing metallicity can be seen. In massive stars of 
decreasing metallicity, it is expected that the amounts of $^{12}$C and $^{32}$S 
would be similar, but the amounts of $^{13}$C and $^{34}$S would decrease (WW95). 

\subsection{New Boundary Conditions}
The observational data used in the models in this paper have been taken 
from RM03, M06, Levashakov et (2006), Anders \& 
Grevesse (1989), Mauersberger et al (2004), Spite et al (2006) and Chavez 
(2008). The data for the carbon and sulphur isotopes are summarised 
in Table~\ref{tab:data}. The ages used have been derived under the assumption 
of the concordant $\Lambda$CDM cosmology.

\subsubsection{Local Constraints}
RM03 have used the solar system value of carbon from Cameron (1982), 
and derive a local ISM value based upon the average of a range of 
observational data; the local ISM value employed in our work is taken
directly from RM03.
Assuming a present-day age of 13~Gyr, the solar system is taken to
have formed at $t=8.5$~Gyr (i.e., 4.5~Gyr ago).  Wherever possible
and relevant, we have used the solar values of Anders \& Grevesse (1989).

For $^{32}$S/$^{34}$S, we use the data compilation from 
Mauersberger et~al. (2004), itself based upon Ding et~al. (2001) for
the solar system value (derived using the Canyon
Diablo triolite) and Chin et~al. (1999) for the local ISM.

Finally, Chavez (2008) has recently measured, for the first time, titanium 
isotope ratios in nearby low-mass stars by 
studying isotopic shifts seen in TiO spectra of M-dwarfs of the halo, 
and thin and thick 
discs. This was done using the 2d-coud\'e spectrograph at the 2.7m 
telescope at 
McDonald Observatory, with a nominal resolving power of $\sim$120k. 
In total, the isotopic ratios for 
11 stars in the metallicity range $-$1$<$[Fe/H]$<$$+$0 have been
derived. The data is shown in Fig. \ref{fig:ti_isotope} and 
tabulated (and fully described) in Chavez (2008).

\subsubsection{Higher Redshift Constraints}
The values obtained recently by M06 for a spiral galaxy at redshift z=0.89 are 
listed in Table~1; 
the average value of the two spiral arm features seen in absorption 
towards a background radio-loud QSO is used. From M06 (figure~1), the mean 
galactocentric distance for these absorption features within the foreground 
galaxy's spiral arms is $\sim$4.5~kpc.

In Levshakov et al (2006), C\,I features associated with a damped Ly$\alpha$ 
system at $z$=1.15 seen in the 
spectrum of HE~0515$-$4414 were analysed 
to derive the $^{12}$C/$^{13}$C ratio. The inferred lower limit of
$^{12}$C/$^{13}$C $>$ 80 suggested, for the first time, that the 
abundance of $^{13}$C in extragalactic clouds was very low.

In order to infer the isotopic abundance patterns of the ISM at the 
time of formation of the Milky Way, we used a sample of 
extremely metal-poor giants ([Fe/H] $\leq$ $-$3) from Spite et~al. (2006).
We employed the ``unmixed'' stars from their sample (i.e., stars which have 
not had their surface abundances affected by deep mixing), and taken the mean
$^{12}$C/$^{13}$C value (19.38) as representative of 
time $t$$\sim$1~Gyr (roughly the timescale of formation for the Galactic
halo).

\begin{figure*}
\includegraphics[width=140mm]{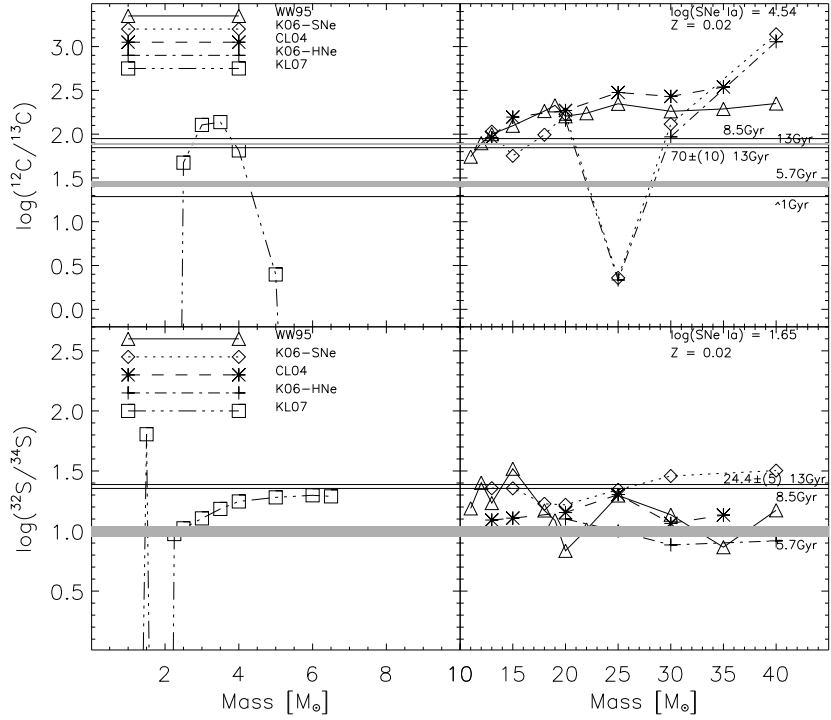}
\caption{Top panel: Carbon isotope ratios as a function of stellar mass at 
solar metallicity for the sets of yields indicated in the inset. 
Bottom panel: As above, but sulphur isotopes. In both panels, the 
Type~Ia supernova (SNe~Ia) values were derived from Nomoto et~al. (1997). 
The grey horizontal lines correspond to the various observational constraints
described in \S~3.3.
\label{fig:c_s.isotopes}}
\end{figure*}

\begin{figure*}
\includegraphics[width=140mm]{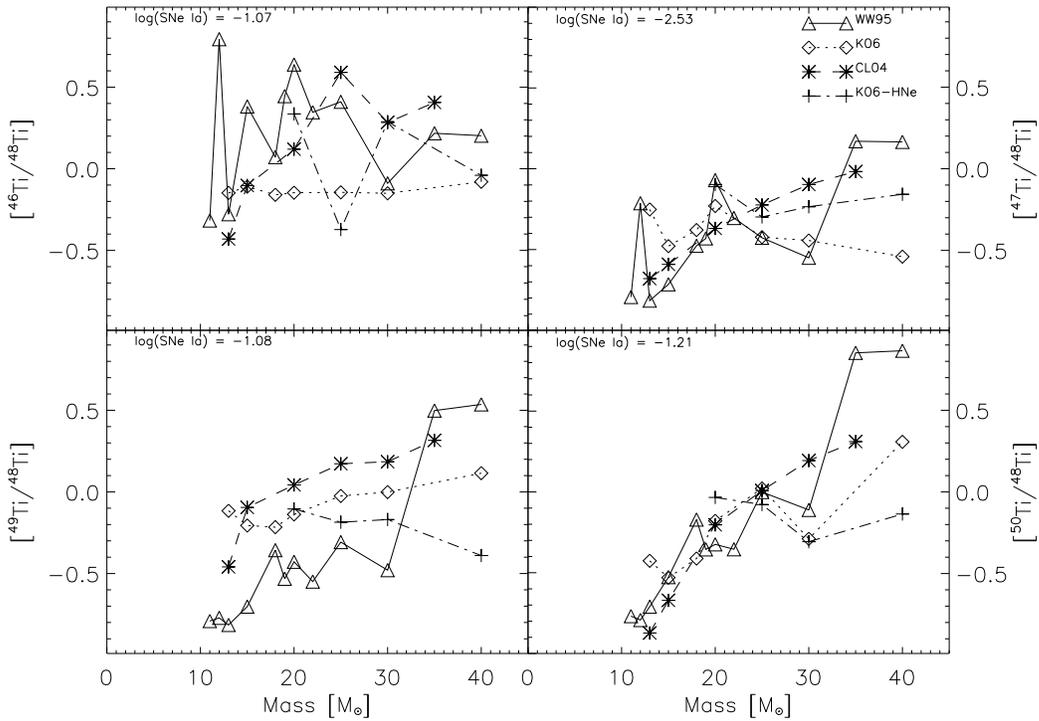}
\caption{Titanium isotope ratios as a function of stellar mass at 
solar metallicity for the sets of yields indicated in the inset.
In all panels, the
Type~Ia supernova (SNe~Ia) values were derived from Nomoto et~al. (1997).
\label{fig:ti_isotopes}}
\end{figure*}

\begin{figure*}
\includegraphics[width=140mm]{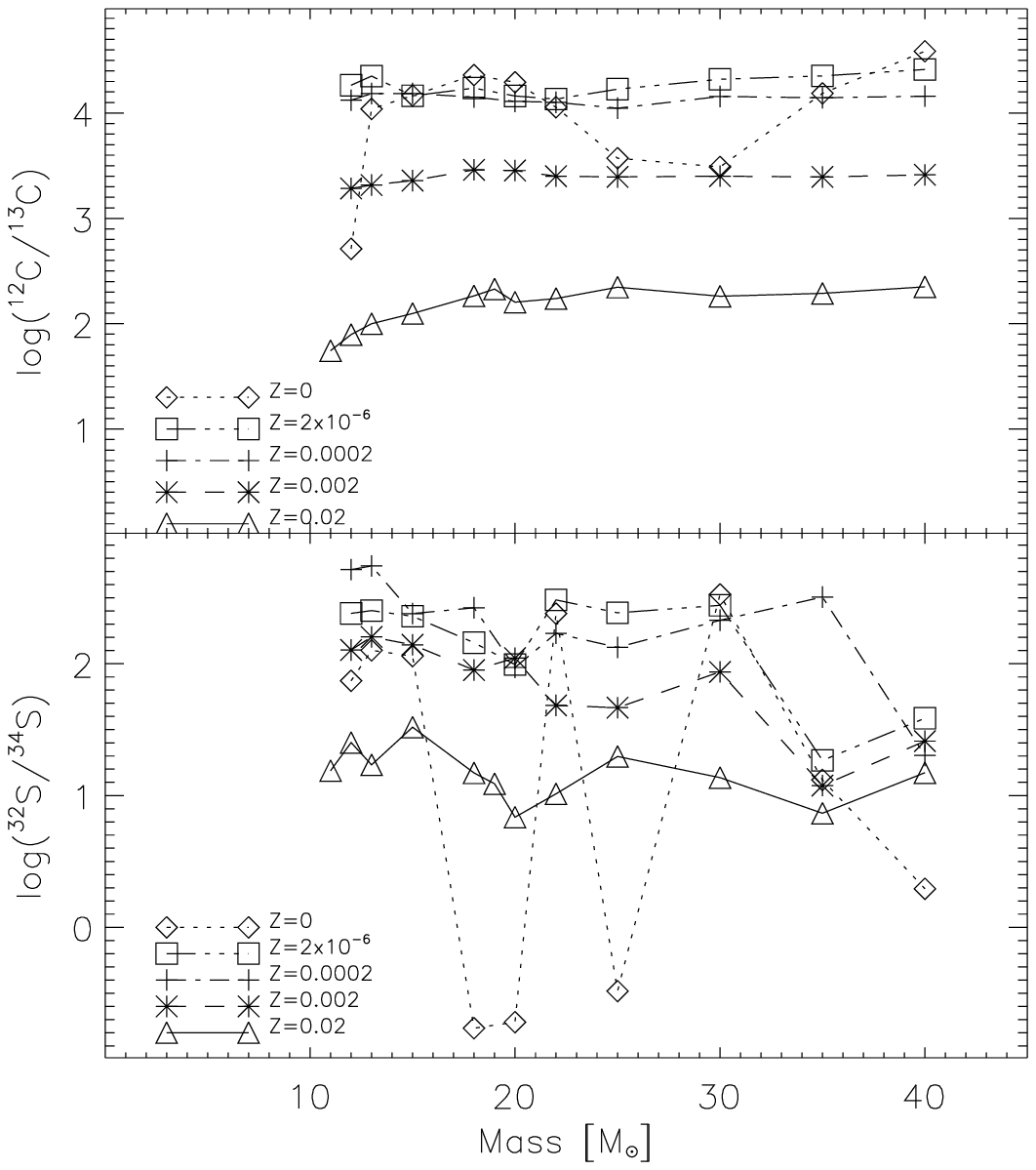}
\caption{Top panel: Carbon (top panel) and sulphur (bottom panel) 
isotope ratios as a function of stellar mass for each of the 
metallicities used by WW95.
\label{fig:ww95_metallicity}}
\end{figure*}

\begin{figure*}
\includegraphics[width=140mm]{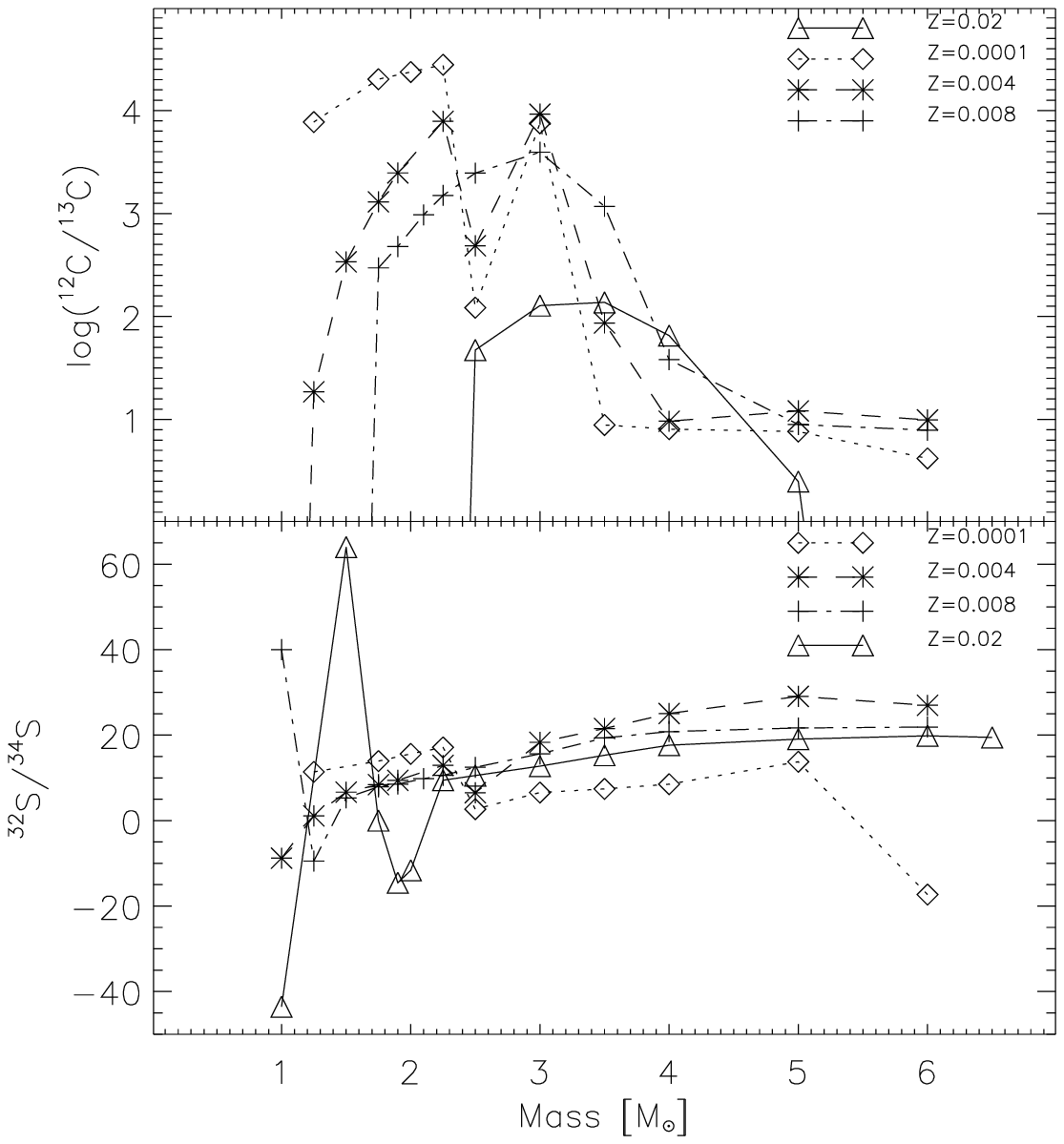}
\caption{Top panel: Carbon (top panel) and sulphur (bottom panel) 
isotope ratios as a function of stellar mass for each of the 
metallicities used by KL07.
\label{fig:kl07_metallicity}}
\end{figure*}

\begin{figure*}
\includegraphics[width=100mm]{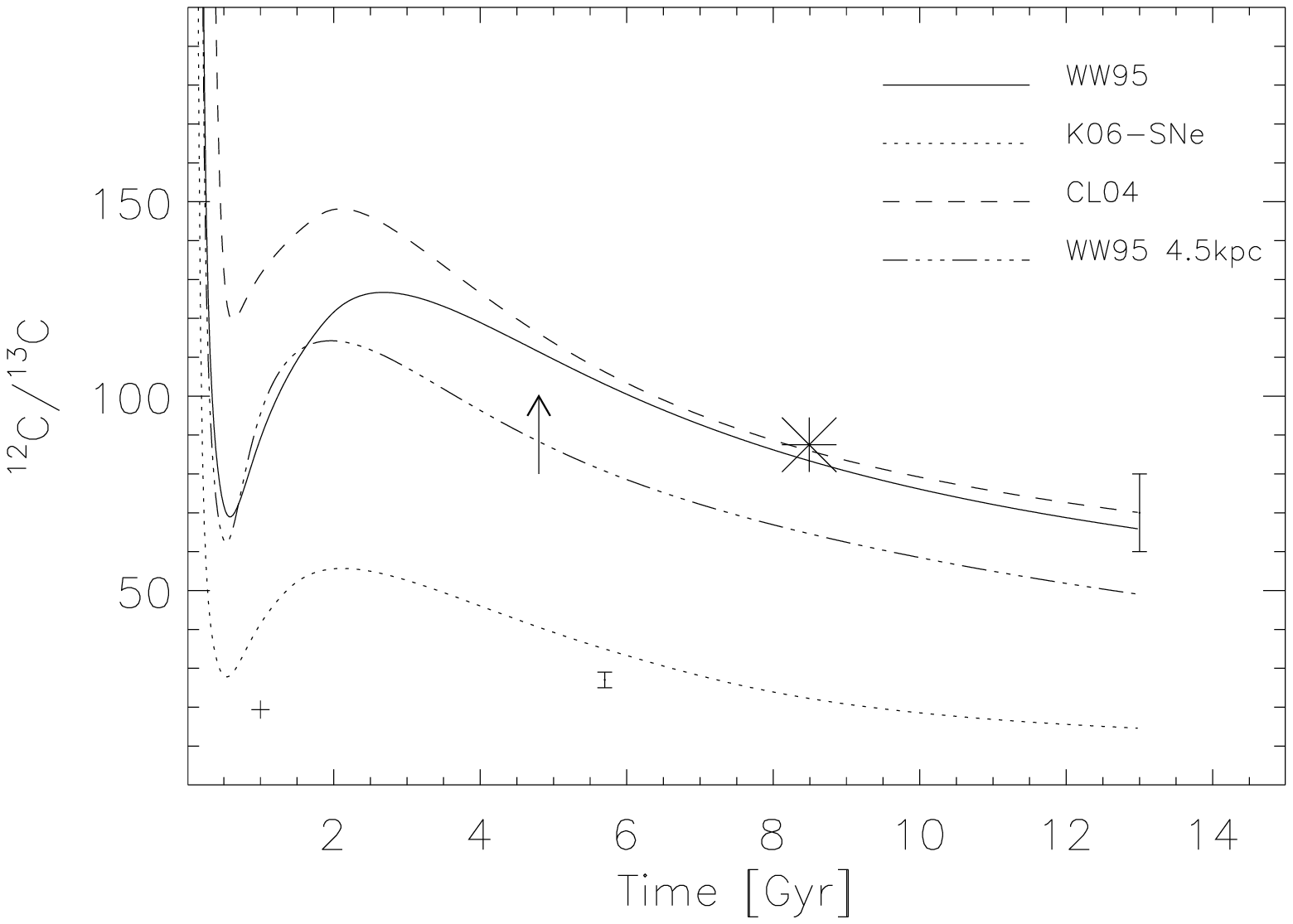}
\includegraphics[width=100mm]{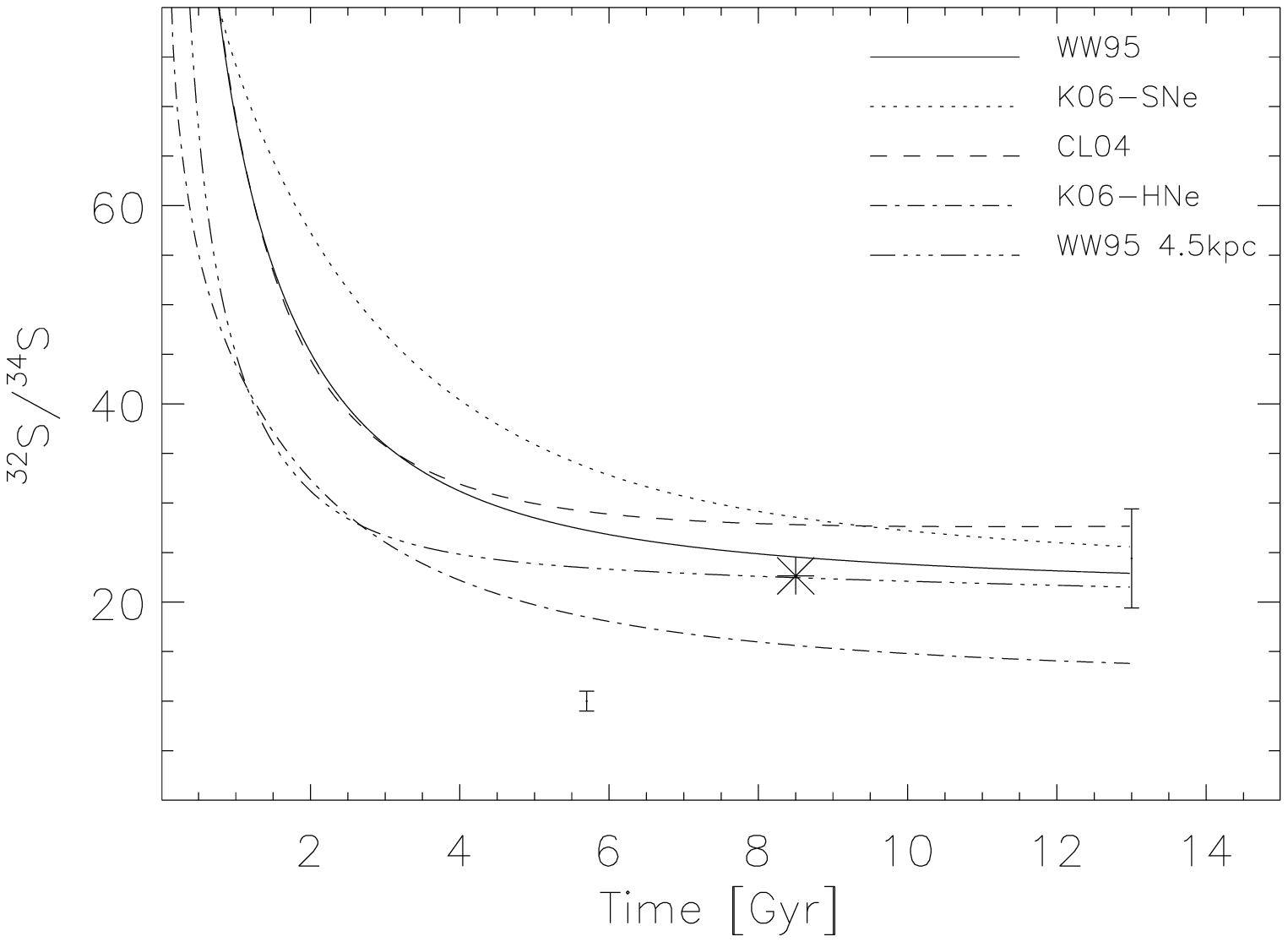}
\caption{Evolution of carbon and sulphur isotope ratios in the 
solar neighbourhood (r = 8.5 kpc) using the KTG IMF for the Type~II supernovae
(SNeII) models of WW95, CL04, and K06. 
For the carbon isotopes: The asterisk, arrow, and plus sign represent the 
solar value from RM03, the high-redshift Levshakov et al (2006) lower limit, 
and the Spite et al (2004) halo star data, respectively; the error bar at 
5.7~Gyr represents the M06 data, and the error bar at 13~Gyr 
correspond to the local ISM values, as reported by RM03.
For the sulphur isotopes: The asterisk represents the solar system value of 
Mauersberger et~al. (2004), while the error bars at 5.7 and 13~Gyr 
correspond to the high-redshift data of M06 and the local ISM 
(Mauersberger et~al. 2004), respectively.
\label{fig:iso_timeevolution}}
\end{figure*}

\begin{figure*}
\includegraphics[width=100mm]{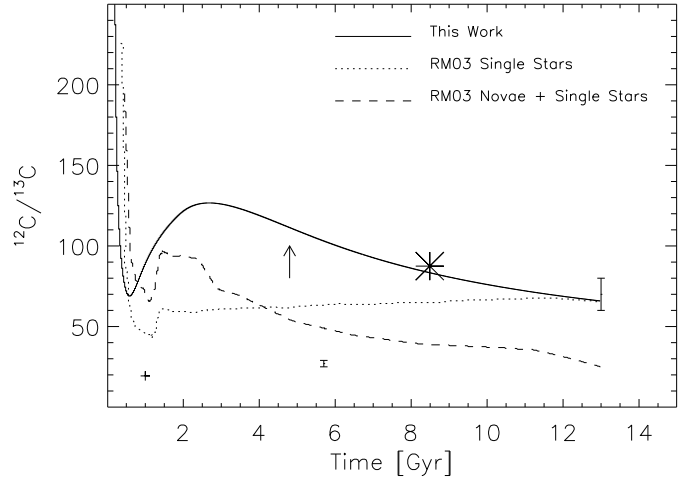}
\caption{Comparison between the evolution of carbon isotope ratios
using the WW95 yields (without novae: solid curve), 
for the template model described here,
alongside the RM03 models both with (dashed curve) and without (dotted curve)
the inclusion of a novae contribution.  An important difference between
the RM03 curves shown here, and those in the corresponding figure~1a of
RM03, is that the models have not been re-scaled \it a posteriori \rm
to match the solar neighbourhood data.  Contemporary low- and intermediate-
mass stellar yields, such as those of KL07, obviate the need for any
such ``re-calibration''.  Symbols are
as in Fig.~\ref{fig:iso_timeevolution}.
\label{fig:rm03}}
\end{figure*}

\begin{figure*}
\includegraphics[width=88mm]{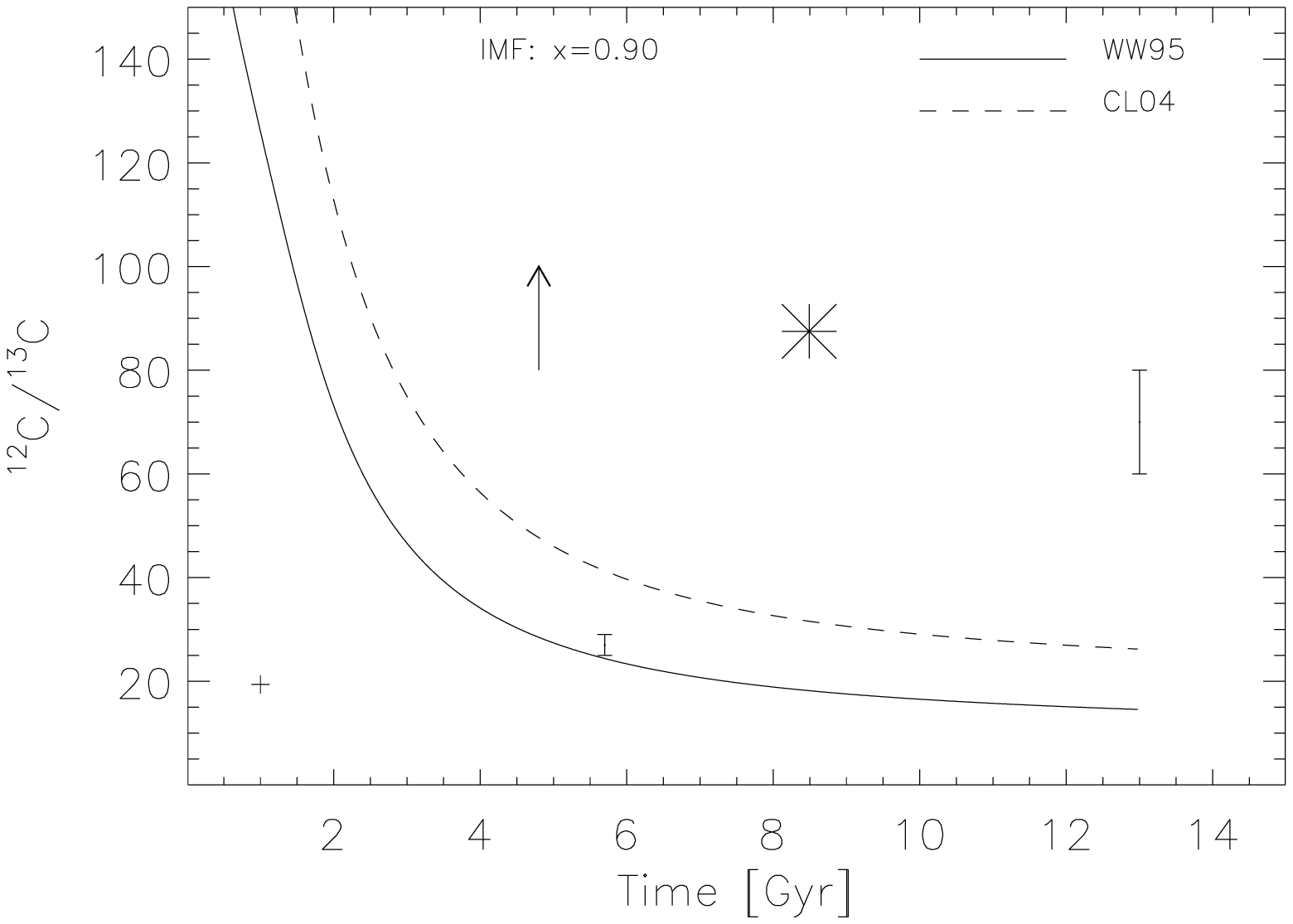}
\includegraphics[width=88mm]{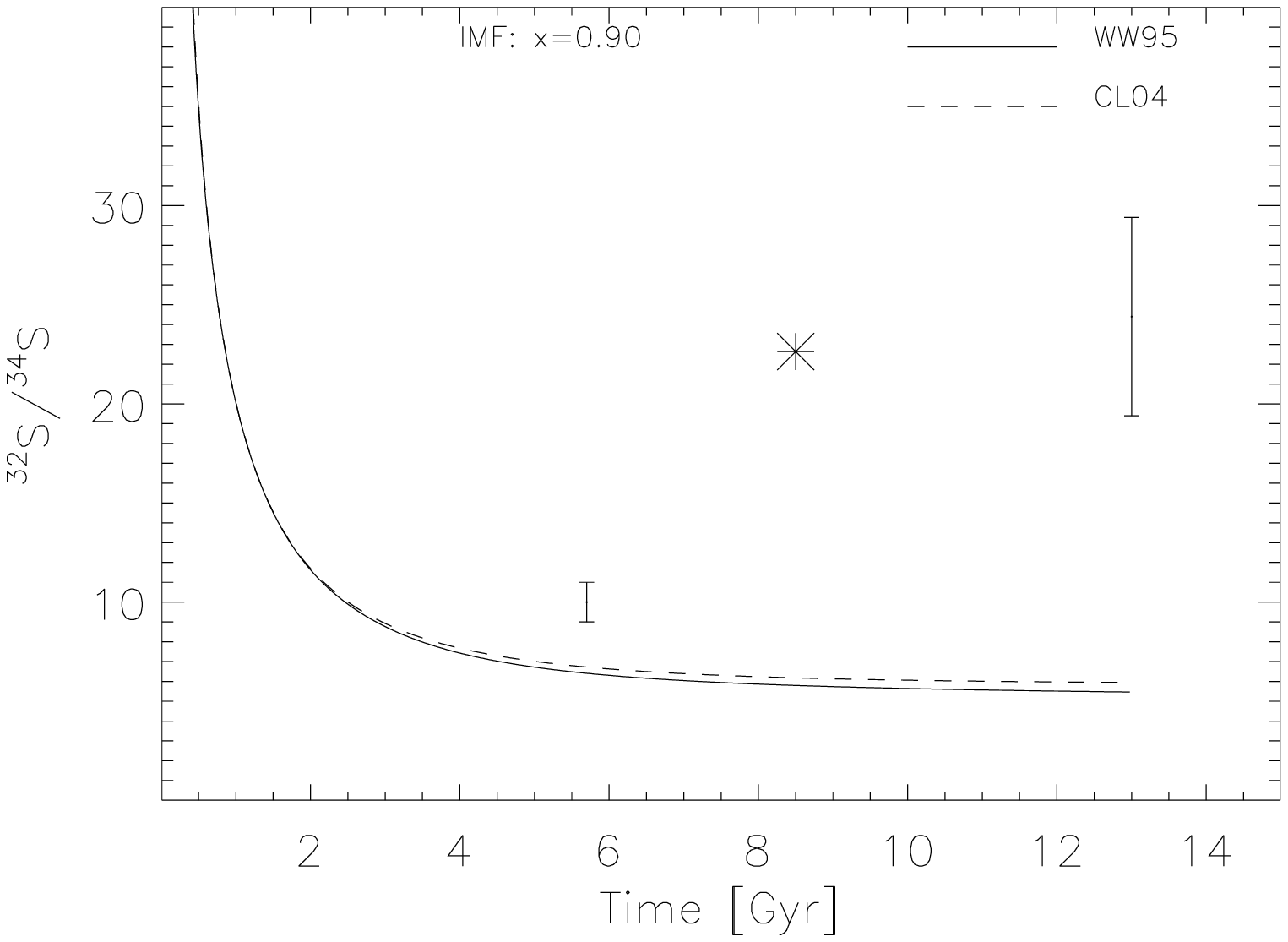}
\includegraphics[width=88mm]{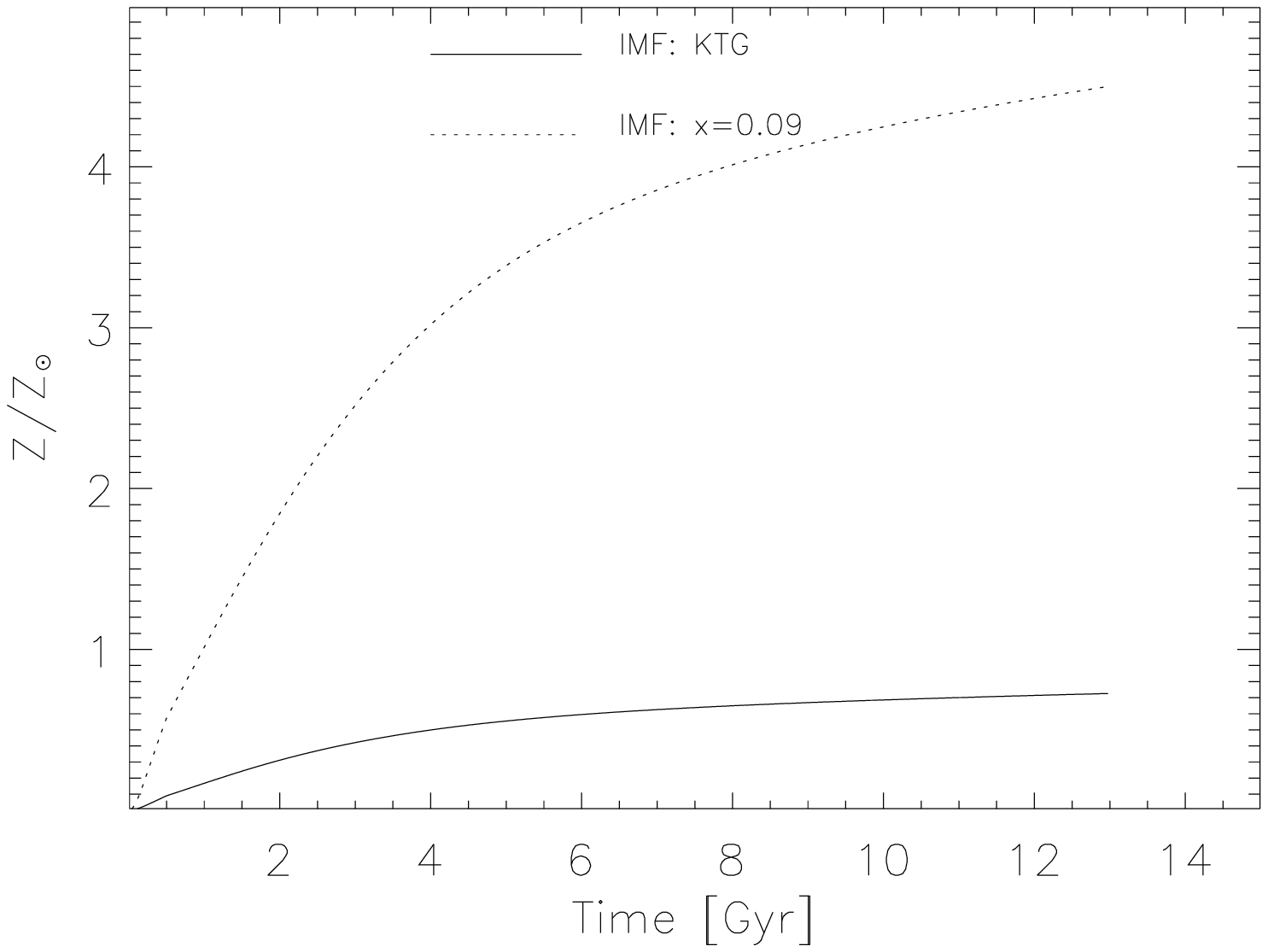}
\caption{The evolution of carbon and sulphur isotope ratios in the 
solar neighbourhood using a single power-law IMF slope of 0.9 
(cf. slope of 1.35 for Salpeter 1955) for the SNeII models of WW95 and 
CL04. The metallicities compared to the solar value for each of the WW95 models
used are also shown. Symbols are as in Fig.~\ref{fig:iso_timeevolution}.
\label{fig:imf}}
\end{figure*}

\begin{figure*}
\includegraphics[width=100mm]{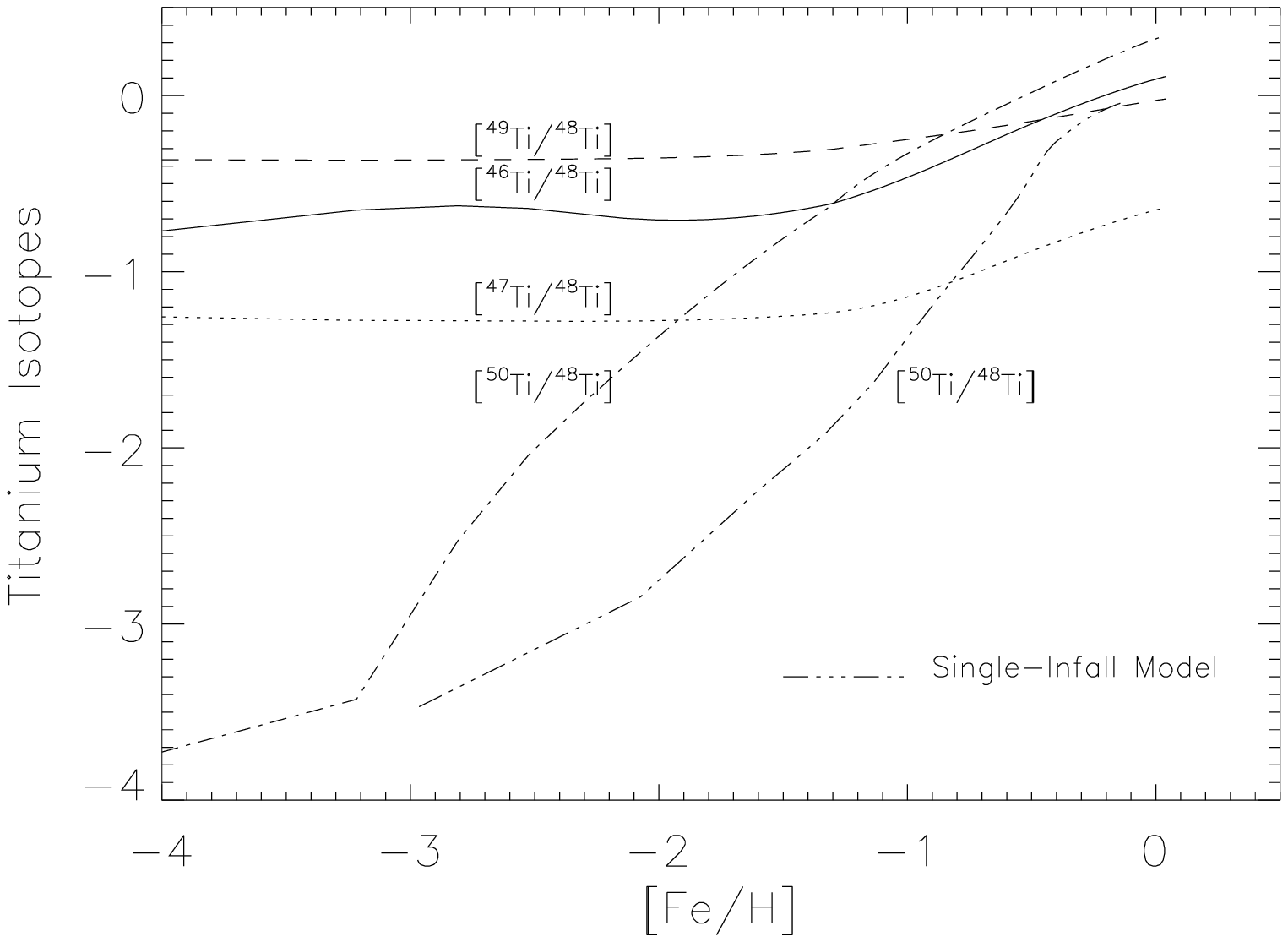}
\includegraphics[width=100mm]{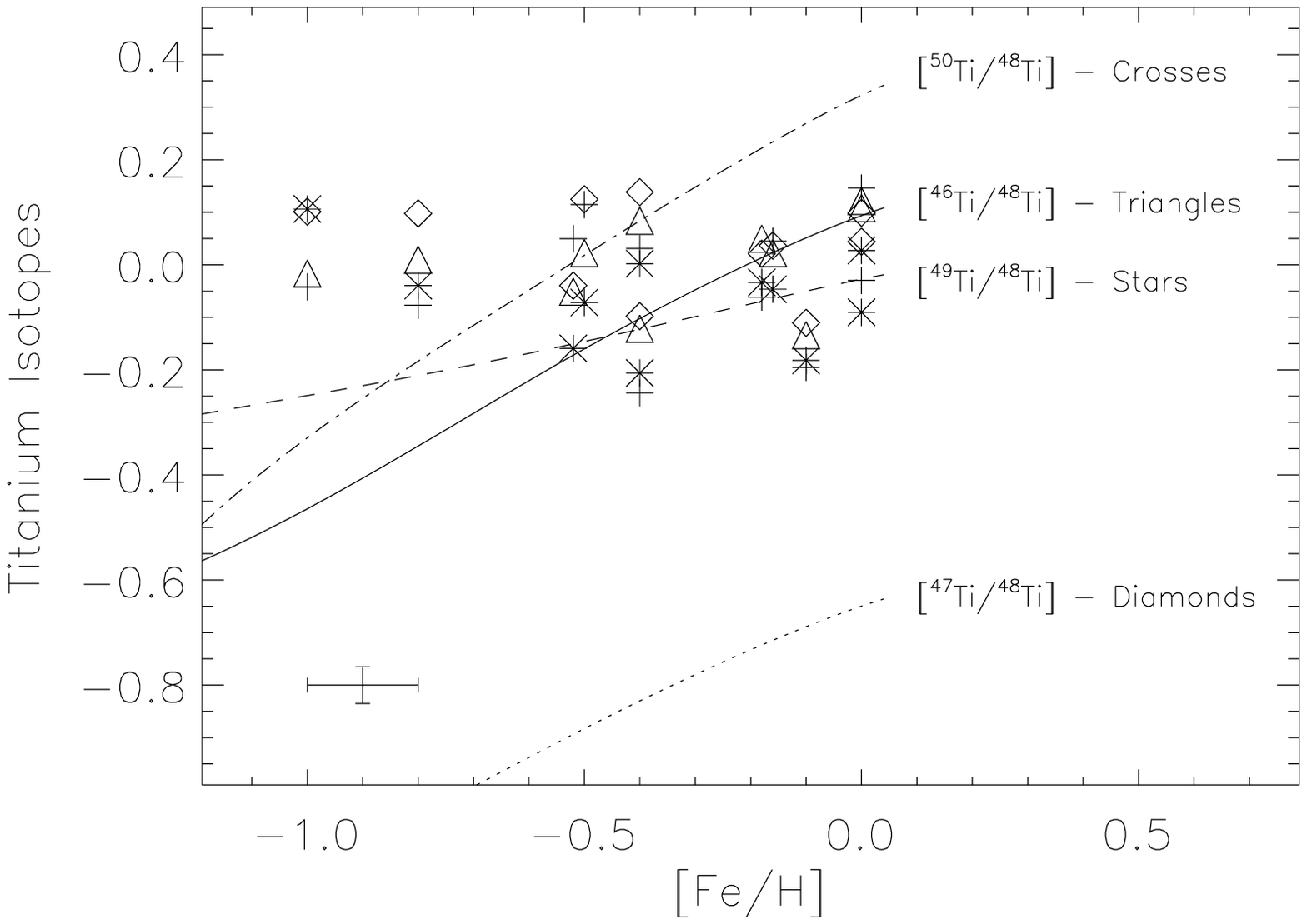}
\caption{The evolution of titanium isotope ratios in the solar 
neighbourhood using the default dual-infall model with the KTG IMF. 
For [$^{50}$Ti/$^{48}$Ti], an older single-infall model (akin to that
of TWW95) is shown for comparison.  In the bottom
panel, the 11 local disc + halo stars observed by Chavez (2008) are 
labelled accordingly. Representative uncertainties are shown in the 
lower left corner 
of the bottom panel; exact star-by-star 
uncertainties are tabulated in Chavez (2008).
\label{fig:ti_isotope} }
\end{figure*}

\section{Results}
Figure~\ref{fig:iso_timeevolution} shows the time evolution of the 
$^{12}$C/$^{13}$C and $^{32}$S/$^{34}$S ratios in our default 
dual-infall model for the solar neighbourhood (\S~3) using three 
different SNeII yield compilations. The observed values for the 
solar system (RM03), local halo stars (Spite et~al. 2006), and the local 
interstellar medium (RM03; Mauersberger et~al. 2004) 
are shown for comparison. The values from M06 are also plotted; as mentioned 
earlier, the mean galactocentric distance probed by the background
QSO towards this high-redshift galaxy is $\sim$4.5~kpc; as such, 
besides the ``solar neighbourhood'' prediction, we have also plotted the 
results of our default model at a galactocentric radius of 4.5~kpc, 
using the WW95 yields.

The evolution of $^{12}$C/$^{13}$C, employing the WW95 and CL04 yields,
is essentially indistinguishable after $\sim$5~Gyrs, while the 
discrepant nature of the models employing the K06 yields is self-obvious.
We have not shown the models using the K06 hypernovae yields, as their 
impact for the evolution of $^{12}$C/$^{13}$C is negligible.  In the 
case of $^{32}$S/$^{34}$S, each of the yield compilations results in 
fairly self-consistent evolutionary trends, although the hypernovae do 
reduce the predicted ratio by a factor of two relative to models
neglecting them.

From Fig.~\ref{fig:iso_timeevolution}, we also note the existence of predicted
positive radial gradients in both 
$^{12}$C/$^{13}$C and $^{32}$S/$^{34}$S, reflecting both the inside-out
galaxy formation framework and the consequent increased importance of 
the secondary production of $^{13}$C in the inner regions of the
galactic model (e.g. RM03).
From the lower panel of Fig.~\ref{fig:iso_timeevolution}, we can see 
that our template 
dual-infall model is consistent with the extant galactic $^{32}$S/$^{34}$S
data (solar system and local ISM); the lower value observed at
z=0.89 (M06) can be partially reconciled with it being nearer its
respective galaxy's centre than the solar neighbourhood,\footnote{Under the 
assumption, of course, that the respective chemical evolution models
for the two systems are comparable; admittedly, an assumption we 
make with little supporting evidence, save for the fact that the 
morphological classifications of these two systems are comparable.}
a point to which we return in \S~4.2.

\subsection{The Need For Novae}
Previous galactic models for $^{12}$C/$^{13}$C (e.g. RM03), 
while consistent with the local ISM, predicted an increase 
in $^{12}$C/$^{13}$C over the past 5-10 Gyrs, driven in part by the 
use of the older van den Hoek \& Groenewegen (1997) yields.  
To ameliorate that apparent discrepancy, RM03 introduced an important 
additional source of $^{13}$C, in the form of novae. While successful in 
recovering the decline in $^{12}$C/$^{13}$C with time, the 
overproduction of $^{13}$C resulted in a significant mismatch between the 
model and the observations, as shown in Fig.~\ref{fig:rm03}, 
which required an 
\it a posteriori \rm re-scaling of the model to the solar system value.

Conversely, the predicted decline in $^{12}$C/$^{13}$C over the past 
5-10 Gyrs in our model is naturally consistent with the observed solar
value and that of the local ISM.  This behaviour is driven by the 
new KL07 yields (which obviously RM03 did not have access to)
without the need of recourse to any additional $^{13}$C novae contribution. 
The putative need for novae might even 
be exacerbated if one were to include, for example, ``born
again'' (i.e., re-ignited) stars such as Sakurai's Object; it has been 
suggested recently that such objects might be the dominant source of 
$^{13}$C in the Universe (Hajduk et~al. 2005).  In the future, a
phenomenological treatment of both novae and such born-again objects
will be implemented within {\tt GEtool}.

\subsection{Isotopes At High-Redshift}
As alluded to earlier, 
it can be seen in Fig.~\ref{fig:iso_timeevolution} that it is difficult for 
our template model to reproduce the isotopic ratios observed in the spiral 
galaxy at z=0.89.  Admittedly, the K06 yields appear to provide a better
fit to $^{12}$C/$^{13}$C at t=5.7~Gyr (upper panel) although, at the same
time, they are less successful for $^{32}$S/$^{34}$S.  We have examined
two possible alternatives to our template solar neighbourhood model
which could explain the lower values observed at high-redshift: (i)
varying the galactocentric distance of the model; (ii) varying the 
IMF.

First, as noted earlier, the data of M06 probes a galactocentric distance
closer to 4.5~kpc, rather than the solar galactocentric distance of 
8.5~kpc; our template Milky Way model for this radial ``bin'' would 
predict a $^{12}$C/$^{13}$C evolution offset by $\sim$20\% from the solar 
neighbourhood value, as shown in Fig.~\ref{fig:iso_timeevolution}. This is, however, 
insufficient to reproduce the observed values.

Second, we explored the dependence of the predicted isotope ratios upon the
relative proportion of massive stars in the IMF by flattening (significantly)
the high-mass end of the IMF (representing the IMF by a single power-law
of slope 0.9).
Figure~\ref{fig:imf} shows the time evolution of the $^{12}$C/$^{13}$C 
and $^{32}$S/$^{34}$S isotope ratios using both this massive-star biased IMF
and the default KTG IMF.  The immediate conclusion to be drawn is that
both ratios decrease dramatically (by a factor of approximately four)
when adopting the massive star-biased IMF.  Having said that, it is 
important to be aware that in large part, this dramatic decrease 
is a consequence of the accelerated \it global \rm 
chemical enrichment produced by the flatter IMF, as these isotopic ratios 
decrease rapidly with increasing metallicity (recall the earlier
discussion surrounding
Figures~\ref{fig:ww95_metallicity} and \ref{fig:kl07_metallicity}), 
as opposed to any mass-dependency in the yields.
While an IMF slope of 0.9 appears to be a viable solution to the
low ratios seen at z=0.89, the predicted global metallicity would
be $\sim$4$\times$ solar; such an extreme value seems unlikely for
a fairly average looking late-type spiral at high-redshift.\footnote{Indeed,
the problem is compounded by having to extrapolate the WW95 and CL04 yields
to such high global metallicities, which exacerbates the rapid
decrease in the isotopic ratios.}

We note that neither scenario appears consistent with the 
low $^{12}$C/$^{13}$C seen in the local halo stars within the Milky Way
(i.e., the t=1 Gyr datum in Fig.~\ref{fig:imf}). To try and reach such low values, 
we examined a range of halo infall timescales and disk infall ``delays'', for 
example, having a 5 Gyr delay between the first and second infall phase. All 
of these attempts to reach the low values seen in halo stars were to no avail, 
however, and the models could not be made to reproduce the observational data. 
This is not completely surprising, as none of the yields used in this work have 
ratios as low as 20, as shown in Fig.~\ref{fig:c_s.isotopes}, other than the 5$M_{\odot}$ value of 
KL07. Therefore, there is no combination of parameters in this work that 
would 
lead to such low values, and the answer must lie in additional 
physics. One such 
solution, and perhaps 
the most likely, is the idea of rotationally-induced mixing at 
low-metallicity, as demonstrated by Chiappini et al (2008), to which we refer 
the reader. 

\subsection{Titanium Isotopes}
Until recently, the chemical evolution of titanium isotopes has been more
of an academic exercise than an experimentally-driven one, in
the sense that observational constraints outside the pre-solar nebula
did not exist (e.g. see the lack of data in TWW95; figure~28).
The classical TWW95 models suggested that all of the titanium 
isotopes were underproduced with respect to the solar values, 
with $^{47}$Ti and $^{50}$Ti being particularly problematic. Two developments 
over the past decade make it timely to revisit this issue: (1) dual-infall 
models such as RM03 and our own have supplanted the simpler 
monolithic-like collapse models of TWW95, and (2) recent 
observational data from Chavez (2008) have, for the first time, provided 
stellar values (outside the pre-solar nebula) against which to confront models.

In Fig.~\ref{fig:ti_isotope} (top panel) we show the predicted behaviour 
of [$^{46,47,49,50}$Ti/$^{48}$Ti] vs [Fe/H] for our template dual-infall model, compared with the single-infall model used to replicate the results found 
by TWW95.\footnote{The primary difference between the 
single- and dual-infall models is manifest in the behaviour of 
$^{50}$Ti/$^{48}$Ti, with the others differing by less than a factor of
two; as such, we only show the former in Fig.~\ref{fig:ti_isotope}.}

Figure~\ref{fig:ti_isotope} (bottom panel) again shows the dual-infall 
model predictions as well as the ratios observed in the 11 disc and halo 
stars from Chavez (2008).  We can identify three interesting conclusions 
from our preliminary analysis of this dataset:

\begin{enumerate}
\item the underproduction of $^{50}$Ti claimed by TWW95 
appears to be, in large part, an artifact of the single-infall model;
\item $^{47}$Ti remains problematic; whether this reflects an important
missing nucleosynthetic source from our chemical evolution models, 
such as helium detonation in sub-Chandrasekhar mass Type~Ia supernovae,
remains unclear;
\item most importantly, 
our models predict a positive correlation between metallicity and 
[$^{46,47,49,50}$Ti/$^{48}$Ti] over the range of metallicity sampled by 
the observations, while the data are more suggestive of a lack of 
correlation (or even a slight anti-correlation). If confirmed, this 
would be very difficult to understand within the context of existing
galactic chemical evolution models.
\end{enumerate}

\section{Summary}
We have explored the evolution of carbon, sulphur, and titanium isotopes in 
both the local- (Milky Way) and high-redshift Universe, using a suite of 
chemical evolution models generated with {\tt GEtool}.  We examined the
need for a novae contribution to explain the carbon isotope patterns in 
the Milky Way, the evidence for a massive star-biased IMF at 
high-redshift based upon sulphur isotope ratios, and the impact of new
observations of titanium isotope patterns in nearby stars upon our
picture of galactic chemical evolution.  We have found:

\begin{itemize}
\item In contrast to earlier studies, the necessity for a significant
contribution of $^{13}$C from novae is ameliorated when employing 
contemporary models of asymptotic giant branch stars.\footnote{It should
be noted that novae, or some additional source, \it will \rm be required
though to explain isotopic ratios such as $^{14}$N/$^{15}$N and 
$^{16}$O/$^{18}$O; the KL07 yields do not assist in this regard.}
\item A massive star-biased IMF at high-redshift results in a significant
decrease in the predicted $^{12}$C/$^{13}$C and $^{32}$S/$^{34}$S, 
consistent with those observed, but at the expense of predicting 
highly super-solar metallicity in otherwise normal-looking 
spiral galaxies.
\item Earlier studies which suggested that $^{50}$Ti was significantly
underproduced were, in part, led to this conclusion by 
the adoption of older monolithic-style
collapse models of galactic evolution.  Our dual-infall model eliminates
these apparent problems, although it remains true that 
$^{47}$Ti is problematic (i.e., underproduced).
\item Our chemical evolution models predict a positive correlation between
trace-to-dominant titanium isotope ratios and metallicity, while the data 
is more suggestive of a lack of correlation.
\end{itemize}

\section*{Acknowledgements}

BKG acknowledges the support of the UK's Science \& Technology
Facilities Council (STFC Grant ST/F002432/1) and the 
Commonwealth Cosmology Initiative. LC acknowledges the support of the 
CONACyT grants 46904 and 60354. PSB acknowledges the support
of a Marie Curie Intra-European Fellowship within the 6th
European Community Framework Programme.
All modeling and analysis was carried out on the University of Central
Lancashire's High Performance Computing Facility.

\label{lastpage}

\end{document}